\documentclass[aip,amsmath,amssymb,reprint]{revtex4-2}
\usepackage{color}
\usepackage{hyperref}
\usepackage{enumitem}
\usepackage{graphicx,subfigure}
\usepackage[normalem]{ulem}
\usepackage{lineno}
\usepackage{booktabs}
\usepackage{balance}
\usepackage{array}
 \usepackage{multirow}
 \usepackage{hhline}
 \usepackage{makecell}
  \usepackage[symbol]{footmisc}
    \usepackage{siunitx} 
\newif\ifHighlitedChanges
\def\ifHighlitedChanges{\iftrue}
\ifHighlitedChanges
   
  \def\STRIKE#1{{\color{red}\sout{#1}}}
\else
  
  \def\STRIKE#1{\relax}
\fi

\begin{document}
\title{Compositional fluctuations and polymorph selection in crystallization of model soft colloids}
\author{Abhilasha Kumari} 
\affiliation{School of Chemical and Materials Sciences, Indian Institute of Technology Goa, Goa, 403401, India}
\author{Gadha Ramesh}
\affiliation{Department of Physics, Indian Institute of Science Education and Research (IISER) Tirupati, Tirupati, Andhra Pradesh, 517619, India}
\author{Debasish Koner}
\affiliation{Department of Chemistry, Indian Institute of Technology, Hyderabad, Telangana, 502285, India}
\author{Rakesh S. Singh}
\email{rssingh@iisertirupati.ac.in}
\affiliation{Department of Chemistry, Indian Institute of Science Education and Research (IISER) Tirupati, Tirupati, Andhra Pradesh 517619, India}
\author{Mantu Santra}
\email{mantu@iitgoa.ac.in}
\affiliation{School of Chemical and Materials Sciences, Indian Institute of Technology Goa, Goa, 403401, India}

\begin{abstract}
Understanding polymorph selection in atomic and molecular systems and its control through thermodynamic conditions and external factors (such as seed characteristics) is fundamental to the design of targeted materials and holds great significance in materials sciences. In this work, using Monte Carlo simulations on the Gaussian Core Model and Hard-Core Yukawa colloidal systems, we investigated the control of polymorph selection and explored the underlying mechanisms by tuning thermodynamic parameters. We demonstrate that by carefully modifying the free energy landscape to render the globally stable face-centered cubic (FCC) phase metastable with respect to the body-centered cubic (BCC) phase, the polymorphic identity of particles transitions from FCC-dominated to BCC-dominated via an intermediate regime where both phases nucleate --- either selectively or competitively --- giving rise to a critical-like composition fluctuation of the growing solid-like cluster during the nucleation process. We further probed the critical solid-like cluster compositions, especially in the vicinity of the triple point where the three phases coexist, and observed an interpenetrating arrangement of FCC- and BCC-like particles rather than a commonly observed non-classical core-shell-like two-step nucleation scenario. In addition, we investigated the polymorph selection signatures encoded in local structural fluctuations of the metastable fluid using a machine learning approach based on structural descriptors derived from persistent homology, a topological data analysis method. We believe that the insights gained from this work have the potential to add to the ongoing efforts to control crystallization pathways to obtain the desired functional material. 
\end{abstract}
\keywords{phase transition, nucleation, polymorph selection, machine learning, composition fluctuations}
\maketitle

\section{\label{sec_1}Introduction}
Crystallization is a fundamental process with wide-ranging applications in various research areas, including materials, geological, and medicinal sciences, as well as numerous industrial applications~\cite{myerson_book, sosso2016}. A key objective in this domain is the design and synthesis of solid structures or a polymorph of functional relevance (\textit{i.e.}, possessing specific, desirable properties)~\cite{predictive_2012, directed_2021} --- a field often referred to as ``crystal engineering.'' This is particularly important in the pharmaceutical industry, where the selection of an appropriate polymorphic form of a drug can significantly influence its efficacy and stability. Crystal engineering relies heavily on achieving a molecular-level understanding of phase transformation pathways and the mechanisms that govern polymorph selection. As a result, there is growing interest in using advanced experimental technique to probe the molecular-scale mechanisms that govern polymorph formation during crystallization~\cite{sleutel_nature_2018, xing_2019, zhou_2019, houben_2019, nakamuro_2021, nozawa_comphys_2025}.

The crystallization mechanism and kinetics of a metastable fluid undergoing transformation into a stable solid phase are often described by classical nucleation theory (CNT)~\cite{becker-doring-1935, frenkel-book-1955, pablo_book}. In its simplest form, CNT assumes that nucleation proceeds via the formation and subsequent growth of an embryo whose structural and thermodynamic properties are identical to those of the final stable solid phase. The energetics of embryo formation are governed by a competition between the free energy gain due to the fluid-to-solid transformation and the interfacial free energy cost of creating a fluid-solid interface. The net free energy change determines the nucleation barrier and thus the kinetics of crystallization. However, many recent experimental and computational studies have revealed nucleation pathways that deviate significantly from the CNT predictions. These observations suggest a more complex, so-called ``non-classical'' nucleation mechanisms~\cite{vekilov2010, sear2012, delhommelle_2011, yoreo_science_2014, sleutel_pnas_2014, peng2015, bagchi_osr_2013, bagchi_ice_2014, poole_1, poole_2, frenkel_jcp_triple, tanaka_natcomm_2022, caroline_2007, dijstra_2015} where nucleation often proceeds via a multi-step pathway --- the first step involves the appearance of local fluctuations or precursors that are structurally distinct from the final stable phase. 

For systems with a complex free energy landscape containing multiple metastable phases, such precursors can resemble an intermediate metastable phase that lies closer in structure or free energy to the parent fluid~\cite{mat_1, mat_2, mat_3, lekkerkerker_2002, chung_nat_phys, bagchi_2dm, bagchi_ss_2013, sciortino_natphys_2014, ishizuka2016, sleutel_nature_2018, van2008}. In other scenarios --- for example, in the crystallization of globular proteins~\cite{frenkel_science, vekilov_2004, zhang2017}, calcium carbonate~\cite{gebauer2008, wallace2013, yoreo_science_2014, demich2011}, and urea~\cite{salvalaglio2015} --- multi-step nucleation is mediated by dense disordered precursor phases. In protein crystallization, this precursor formation is associated with a submerged (metastable) gas-liquid criticality, which leads to an enhancement of density fluctuations in the metastable system, and in turn, to a significant alteration of the nucleation pathway and kinetics~\cite{frenkel_science}. There is also now growing interest in uncovering early-stage nucleation signatures that may dictate polymorph selection from a metastable fluid~\cite{tan2014visualizing, kawasaki2010formation, sleutel_nature_2018, xing_2019, zhou_2019, houben_2019, nakamuro_2021, marjolin_2023}.

The above observations that challenge the CNT-based picture underscore the need for a deeper understanding of crystallization pathways, particularly during the early stages of nucleation, where particles begin to locally order within the metastable fluid. These early-stage structural fluctuations can be crucial in determining the crystallization path or eventual selected crystalline polymorph. Furthermore, gaining control over the phase transition pathway by manipulating the underlying free-energy landscape is of significant interest, especially in the vicinity of a triple point, where three phases coexist and the indirect involvement of an intermediate phase in the transition cannot be neglected. In terms of theoretical understanding, the emergence of non-classical nucleation pathways has often been rationalized using phenomenological theoretical approaches, such as, rooted in the CNT within capillary approximations~\cite{iwamatsu_2011, iwamatsu_2012_2, demo_2020} and kinetic approaches based on diffusive growth~\cite{iwamatsu_2012, iwamatsu_2017}, in addition to the approaches based on the classical density functional theory (c-DFT).~\cite{oxtoby_t1, oxtoby_t2, bagchi_osr_2013, guo2016, bagchi_ice_2014, puja_1, puja_2, yuvraj_2023, yuvraj_2025} Furthermore, computational studies employing minimal lattice models~\cite{whitelam_2010, whitelam_2011, poole_1, poole_2} have provided valuable insight into the kinetics and mechanism of such transitions. Nevertheless, in off-lattice molecular and atomic systems that go beyond hard-sphere models, a comprehensive understanding of polymorph selection pathways and the subtle, often hidden signatures encoded in the metastable fluid remains elusive. In off-lattice models, along with the possibility of diverse local structural fluctuations in the metastable fluid, particle transport, can also become an important factor, to dictate the nucleation mechanism. Therefore, developing an understanding of nucleation in off-lattice systems is essential for advancing control over crystallization in more realistic systems relevant to real-world applications.  

To gain fundamental insights into crystallization mechanisms at the molecular level in real-world systems, colloidal suspensions are frequently employed as model systems because their mesoscopic size enables single-particle resolution in experiments. In this work, we employ \textit{in silico} models of colloidal systems --- specifically, the Gaussian Core Model (GCM)~\cite{gcm} and the Hard-Core Yukawa (HCY) model~\cite{meijer_1997} --- which represent two distinct classes of colloidal interactions ranging from hard to soft core interactions. These model systems are well suited for such investigations, as they exhibit a rich phase behavior, including fluid–FCC (face-centered cubic), fluid–BCC (body-centered cubic), and FCC-BCC co-existence lines~\cite{santi_gcm_2005, tanaka_soft_2012, meijer_1997, kratzer_softmater_2015}, which merge at a triple point (the HCY system exhibits two such triple points). Our aim is to develop a molecular-level understanding of the pathways leading to the formation of different crystal polymorphs and the underlying mechanisms governing polymorph selection, especially in the vicinity of the triple point where three phases (fluid, BCC and FCC) coexist. In addition, we have investigated how structural fluctuations in the metastable fluid encode information about polymorph selection. 

Using Monte Carlo (MC) simulations~\cite{frenkel_2005} on the aforementioned atomistic models, we demonstrate that careful modulation of the free energy landscape (specifically, altering the stability of the FCC and BCC phases) leads to a transition in the polymorphic identity of the nucleating particles. This transition proceeds from FCC-dominated to BCC-dominated regimes, via an intermediate region (in the vicinity of the triple point), where both phases nucleate either selectively or competitively, resulting in critical-like composition fluctuations. We further examined the composition of critical solid-like clusters, particularly near the triple point, and observed an interpenetrating arrangement of FCC and BCC-like particles --- distinct from the expected non-classical core-shell-like (or, wetting-mediated) two-step nucleation mechanism~\cite{iwamatsu_2011, bagchi_osr_2013, poole_1}. Furthermore, we also investigated polymorph selection signatures encoded in local structural fluctuations of the metastable fluid using a machine learning approach built on structural descriptors derived from persistent homology, a topological data analysis method. 

The rest of this paper is organized as follows. Section~\ref{sec_2} details the computational protocol followed for simulations of the soft colloidal system modeled via the GCM~\cite{gcm} and HCY~\cite{meijer_1997} potentials. In Section~\ref{subsec_1} we present the phase transition pathways (Section~\ref{subsubsec_1}) and the composition of the critical solid-like cluster, including the spatial distribution of particles with different polymorphic identities within the critical cluster (Section~\ref{subsubsec_2}). In Section~\ref{subsec_3} we present our machine learning scheme to infer hidden information about polymorph selection in metastable fluid structures, and Section~\ref{sec_4} summarizes the major conclusions from this work. 
 \section{\label{sec_2}Model and Simulation Details}
 \subsection{Model details}
\begin{figure}[t!]
    \centering
    \includegraphics[width=0.8\linewidth]{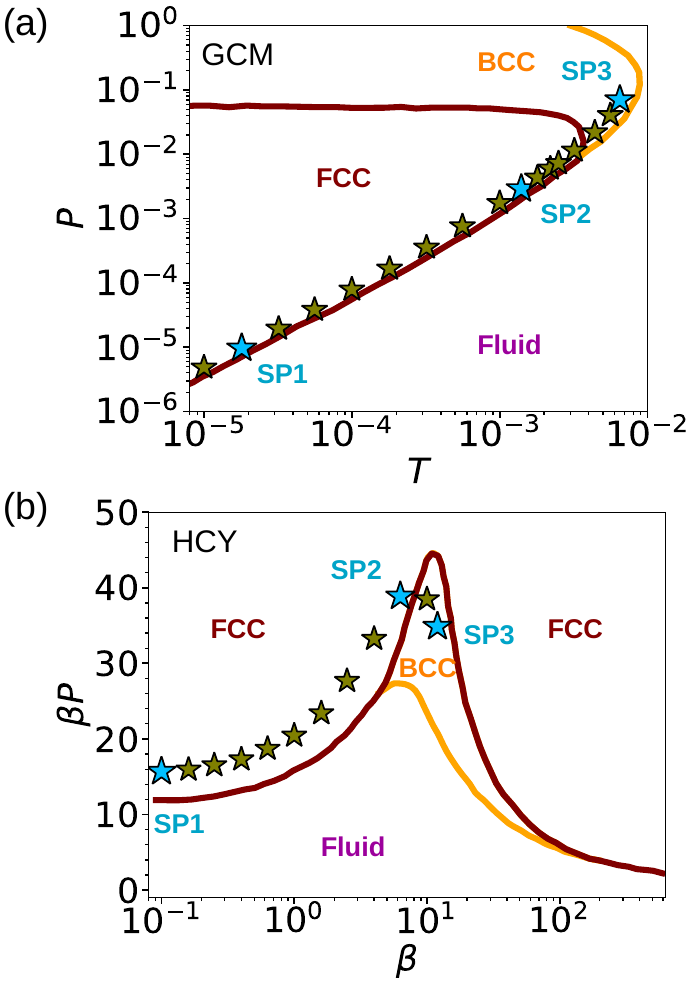}
    \caption{Phase diagram of the GCM and HCY model systems showing the FCC-fluid, BCC-fluid and FCC-BCC coexistence lines. The thermodynamic state points studied in this work (see Table~\ref{tab1}) are marked with $\star$. We have chosen three representative state points: SP1, SP2 and SP3, for both the models systems. At SP1, the FCC is the globally stable phase, at SP2, the FCC phase is globally stable but only marginally stable compared to the BCC phase, and at SP3, the BCC is the globally stable phase.}
    \label{fig1}
\end{figure}

We performed Monte Carlo (MC) simulations~\cite{frenkel_book} in the isothermal isobaric ($NPT$) ensemble on systems interacting via HCY and GCM potentials under the thermodynamic conditions reported in Table~\ref{tab1}. The HCY potential is given as,  
\begin{equation}\label{eq:potential}
   u_{\rm HCY}(r) = \left\{
    \begin{array}{lc}
      \infty, &  r \le \sigma  \\
        \epsilon \frac{\exp \left[-k(r/\sigma - 1)\right]}{r/\sigma}, &   ~~r > \sigma {\rm \;\; }
     \end{array}
   \right.\;,
\end{equation}                                       		
where $\sigma$ is the hard-core diameter, $\epsilon$ is the energy at contact distance, and $k$ is the inverse Debye screening length. In this study, we truncate the HCY potential ($u_{\rm HCY}(r)$) at an inter-particle distance ($r$) of $3.0\sigma$ and shifted it to zero. We chose $k\sigma = 5$, a value commonly used in previous studies of HCY systems, as the corresponding phase diagram is well suited for our purpose~\cite{meijer_1997, kratzer_softmater_2015}. The GCM potential is defined as, $u_{\rm GCM}(r) = \epsilon \exp [-\left(r / \sigma\right)^2]$~\cite{gcm}, where $\epsilon$ denotes the energy at zero inter-particle separation. This potential is truncated at a distance $4.0\sigma$ and shifted to zero. Here, $\sigma$ and $\epsilon$ serve as units of length and energy, respectively. The pressure–temperature ($P–T$) phase diagrams of both model systems are well-documented~\cite{tanaka_soft_2012, santi_gcm_2005}, and are shown in Fig.~\ref{fig1}. Both systems exhibit rich phase behavior, including two (FCC and BCC) polymorphic solid phases that highlight the complexity of the system’s free energy landscapes.

For both model systems, we first selected a state point in the region of the phase diagram where the FCC phase is the globally stable phase (\textit{i.e.}, at low $T$ for GCM and low $\beta$ for HCY). We then gradually moved towards the region in the $P-T$ plane where the BCC phase becomes globally stable. Along this path, we chose a set of state points such that the fluid remains metastable with respect to the solid (FCC and/or BCC) phase(s), and the size of the critical solid-like cluster ($n^*_{\rm sol}$) --- estimated from the nucleation free energy profile of a solid-like cluster --- is approximately $300$ (see Fig.~S1 in the Supplementary Material). We followed the protocol outlined in Refs.~\emph{\citenum{bagchi_pre_2018, gadha_jcp_2025}} (see Section~IID in Ref.~\emph{\citenum{gadha_jcp_2025}} and Section~IIC in Ref.~\emph{\citenum{bagchi_pre_2018}}) to identify solid-like particles in the metastable fluid and assign their polymorphic identity as FCC or BCC-like. In this work, we have not distinguished between the FCC- and HCP-like particles. The system consists of $N = 6912$ particles in a cubic box for both model systems.
\begin{table}
\caption{\label{tab1} The thermodynamic state points for the HCY and GCM models chosen in this study. At these state points the size of the critical solid-like cluster ($n^*_{\rm sol}$) is approximately $300$.}
\begin{ruledtabular}
\begin{tabular}{cc|cc}
&   HCY &   GCM \\
\hline
\hspace{0.5cm} $\beta$ & \hspace{0.5cm} $\beta P \sigma^3$ & \hspace{0.2cm} $T^*$ & \hspace{0.2cm} $P \sigma^3$ \\
\hline
$0.10$  & $15.695$ & $1.0\times10^{-5}$ & $4.850\times10^{-6}$ \\
$0.16$  & $15.950$ & $1.8\times10^{-5}$ & $9.700\times10^{-6}$   \\
$0.25$  & $16.530$ & $3.2\times10^{-5}$ & $1.950\times10^{-5}$   \\
$0.40$  & $17.300$ & $5.6\times10^{-5}$ & $3.830\times10^{-5}$  \\  
$0.63$  & $18.715$ & $1.0\times10^{-4}$ & $7.900\times10^{-5}$  \\
$1.00$  & $20.450$ & $1.8\times10^{-4}$ & $1.670\times10^{-4}$  \\
$1.60$  & $23.410$ & $3.2\times10^{-4}$ & $3.560\times10^{-4}$  \\
$2.50$  & $27.680$ & $5.6\times10^{-4}$ & $7.620\times10^{-4}$  \\  
$4.00$  & $33.250$ & $1.0\times10^{-3}$ & $1.750\times10^{-3}$ \\ 
$6.30$  & $38.880$ & $1.4\times10^{-3}$ & $2.900\times10^{-3}$  \\
$10.0$  & $38.485$ & $1.8\times10^{-3}$ & $4.330\times10^{-3}$  \\
$12.0$ & $34.870$ & $2.2\times10^{-3}$ & $6.000\times10^{-3}$  \\
 &  & $2.5\times10^{-3}$ & $7.270\times10^{-3}$  \\
  &  & $3.2\times10^{-3}$ & $1.145\times10^{-2}$ \\
 &   & $4.4\times10^{-3}$ & $2.190\times10^{-2}$  \\
  &   & $5.6\times10^{-3}$ & $4.060\times10^{-2}$  \\  
  &   & $6.5\times10^{-3}$ &$7.000\times10^{-2}$  \\  
\end{tabular}
\end{ruledtabular}
\end{table}
The nucleation free energy was computed using umbrella sampling method. The details can be found in Ref.~\emph{\citenum{bagchi_pre_2018}}. To compute the chemical potential difference between the metastable fluid and stable BCC and FCC solid phases, we followed the free energy-based method outlined in Ref.~\emph{\citenum{Verrocchio_2012}}.

\subsection{Computation of the phase transition pathways}
The crystallization nucleation barrier at the state points reported in Table~\ref{tab1} is $\sim 20-30k_{\rm B}T$, and therefore it is not possible to study the phase transition process in computer simulation without using advanced sampling methods. Furthermore, in this study, we are not interested in the temporal events or kinetics of crystallization; rather, we are interested in the selection of a solid polymorph and the structure of the critical solid-like clusters. Therefore, we followed a method in which the nucleation time is reduced and is within the range of computer simulations without altering the nucleation pathway and solid polymorph selection. 

We first computed the nucleation free energy barrier as a function of the largest solid-like cluster using the umbrella sampling method (see Fig.~S1 in the Supplementary Material). After obtaining the free energy profile, we performed a set of independent MC simulations ($\sim200$) each of which starts with different metastable liquid configurations. In these simulations, the free energy profile obtained from the umbrella sampling is first inverted, \textit{i.e.}, $U_{\rm b}(n^{\rm lc}_{\rm sol}) = -G(n^{\rm lc}_{\rm sol})$, where $U_{\rm b}$ is the biasing potential, $G$ is the free energy obtained from the umbrella sampling and $n^{\rm lc}_{\rm sol}$ is the size of the largest solid-like cluster. Next, this potential is applied to the system as a fixed bias throughout the simulation. This is similar to the bias added in Metadynamics simulations, except that the bias potential here is applied at the beginning of the simulation and no further bias is added during the simulation. The potential bias range is in the window of $0 \leq n^{\rm lc}_{\rm sol}\leq n^*_{\rm sol}$. There was no additional bias added to the system during the simulation. Thus, it is an equilibrium run on a modified fixed \textit{effective} potential energy surface. As a result of this biasing, the system experiences a flat free energy surface in the pre-critical cluster region, enabling spontaneous crystallization within the simulation time scale. It is important to note that the bias is applied along the size of the largest solid-like cluster and not its structure; hence, it is not expected to affect the natural selection of polymorphs --- one of the key objectives of this study. We have collected data from $100$ independent successful trajectories and performed detailed analysis (see Section~\ref{subsec_1}). A trajectory is considered to be successful if $n^{\rm lc}_{\rm sol}$ grows beyond the size of $1500$ particles, which is approximately $5$ times of the critical cluster size.  
\begin{figure*}
    \centering
    \includegraphics[width=0.95\linewidth]{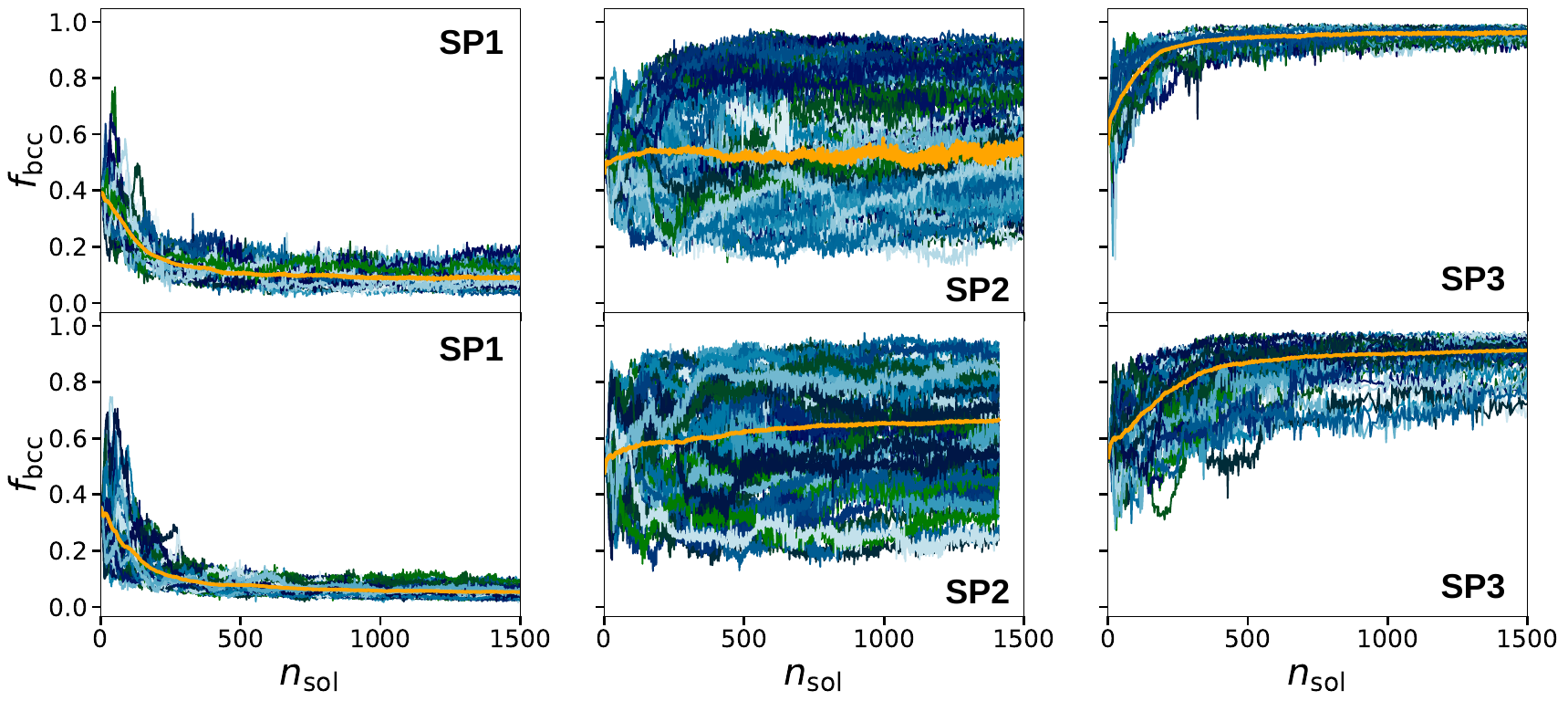}
    \caption{The variation of the fraction of the BCC-like particles in the system ($f_{\rm bcc}$; $f_{\rm fcc}$ is $1-f_{\rm bcc}$) as a function of the number of solid-like particles ($n_{\rm sol}$) during the phase transition for the GCM (top) and HCY (bottom) systems at SP1, SP2 and SP3. We note a crossover in the polymorph selection from FCC to BCC-like via a critical-like enhanced composition fluctuations at SP2.}
    \label{fig2}
\end{figure*}
\begin{figure}
    \centering
    \includegraphics[width=\linewidth]{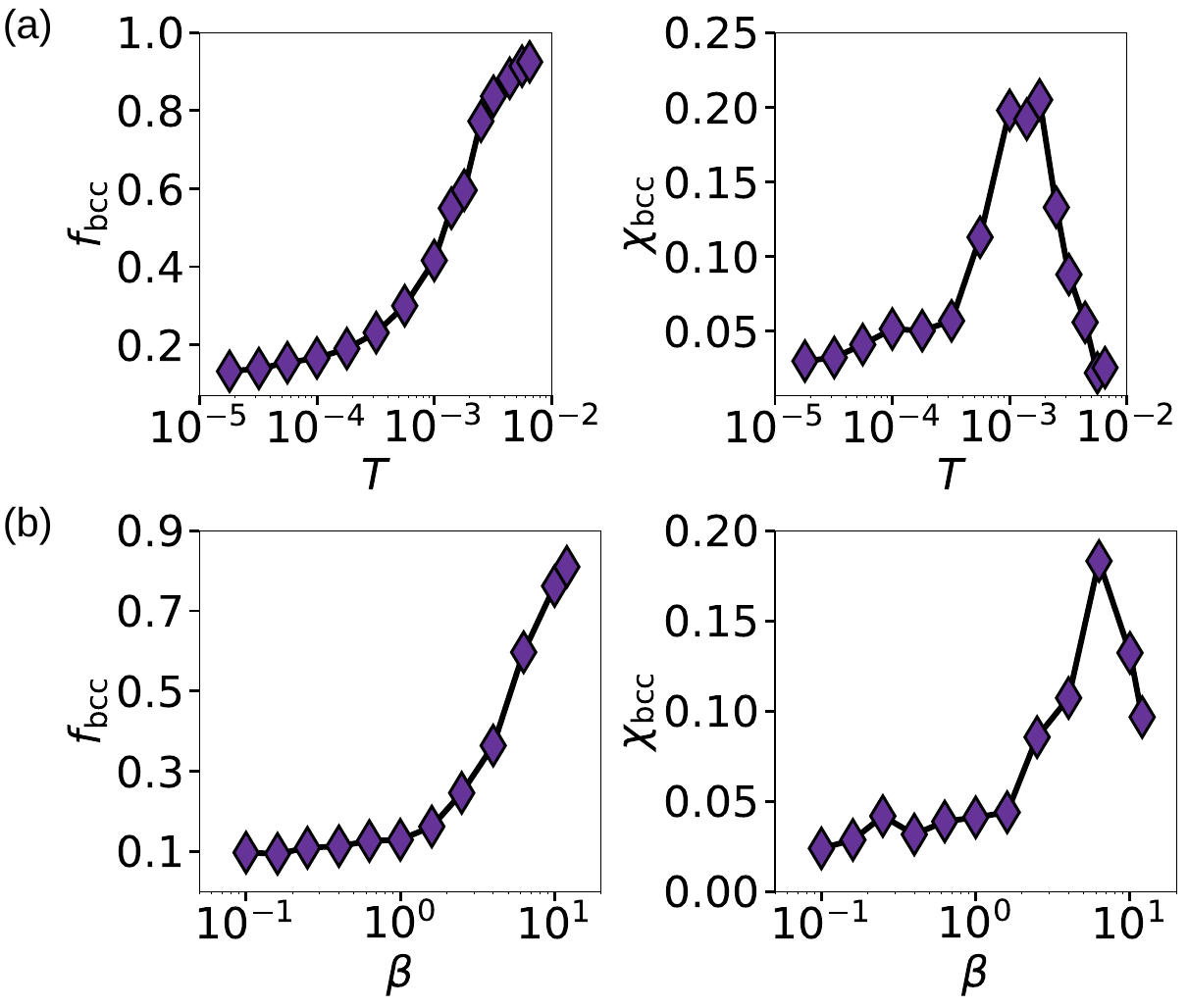}
    \caption{The average fraction of the BCC-like particles ($\langle f_{\rm bcc} \rangle$) along with its fluctuation ($\chi_{\rm bcc}$, defined as $\chi_{\rm bcc} = \langle f_{\rm bcc}^2 \rangle - \langle f_{\rm bcc} \rangle^2$) in the critical solid-like cluster at temperatures corresponding to different state points (see Table~\ref{tab1}) for the GCM (a) and the HCY (b) systems. Here, $\langle ... \rangle$ indicates average over $100$ independent MC trajectories undergoing phase transition.}
    \label{fig3}
\end{figure}
\section{Results and Discussion}
\subsection{Compositional fluctuations and polymorph selection} \label{subsec_1}
\subsubsection{Phase transition pathways and composition fluctuations} \label{subsubsec_1}
We first investigated how the polymorphic identity of solid-like particles and the composition of growing solid-like clusters vary during crystallization in the metastable fluid phase under different thermodynamic conditions ($T,P$) that alter the underlying free energy surface (see Table~\ref{tab1}). As discussed in Section~\ref{sec_2}, the selected state points span a range of scenarios --- from FCC being the globally stable to BCC the globally stable phase. Figure~\ref{fig2} shows the evolution of the polymorphic identity (BCC- or FCC-like) of solid-like particles during crystallization at three representative state points: SP1, SP2, and SP3 (see Fig.~\ref{fig1}). At SP1, FCC is the globally stable phase; at SP2, the FCC phase is only marginally more stable than the BCC; and at SP3, the BCC is the stable phase with respect to the metastable fluid, for both model systems. The chemical potential differences between the metastable fluid and the BCC/FCC phases at SP2 are $\Delta\mu_\mathrm{bcc} = -0.220$ and $\Delta\mu_\mathrm{fcc} = -0.225$ for the HCY model, and $\Delta\mu_\mathrm{bcc} = -0.220$ and $\Delta\mu_\mathrm{fcc} = -0.288$ for the GCM model. Here (in Fig.~\ref{fig2}), we have plotted the fraction of BCC-like particles ($f_{\rm bcc}$) as a function of the total number of solid-like particles ($n_{\rm sol}$) in the system. Note, $f_{\rm fcc} = 1 - f_{\rm bcc}$, as we have not made a distinction between the HCP- and FCC-like solids. As expected, the system predominantly crystallizes into the FCC phase at SP1 and into the BCC phase at SP3, consistent with the respective global thermodynamic stability of these polymorphs at the two state points. However, at SP2, we observe a critical-like scenario in the composition space, for both model systems, where different independent trajectories either prefer BCC-like or FCC-like phase, and sometimes the two (BCC and FCC) phases also grow simultaneously and competitively, giving rise to a varying composition of the growing solid-like cluster. This enhanced polymorph composition fluctuations suggest a critical-like flattening of the underlying free energy surface in the two-dimensional ($f_{\rm bcc}$ and $n_{\rm sol}$) order parameter space (see Fig.~S2 in the Supplementary Material). 

To further characterize the phase transition pathways and fluctuations in polymorph composition on modulating the underlying FES through ($T,P$), in Fig.~\ref{fig3}, we report the variation of the average fraction of BCC-like particles $\langle f_{\rm bcc} \rangle$, along with its susceptibility, defined as $\chi_{\rm bcc} = \langle f_{\rm bcc}^2 \rangle - \langle f_{\rm bcc} \rangle^2$, for both models. On moving from SP1 to SP3, the $f_{\rm bcc}$ increases monotonically, and the $\chi_{\rm bcc}$ shows a maximum at a state point in the vicinity of SP2 where the crossover in the selection of the polymorphic identity of particles occurs for both the model systems. These results unambiguously suggest that the crossover in the selection of the polymorphic identity occurs via an intervening state point region where the FCC/BCC composition fluctuations are greatly enhanced.
\subsubsection{Microscopic structure of the critical solid-like cluster} \label{subsubsec_2}
\begin{figure}[t!]
    \centering
    \includegraphics[width=\linewidth]{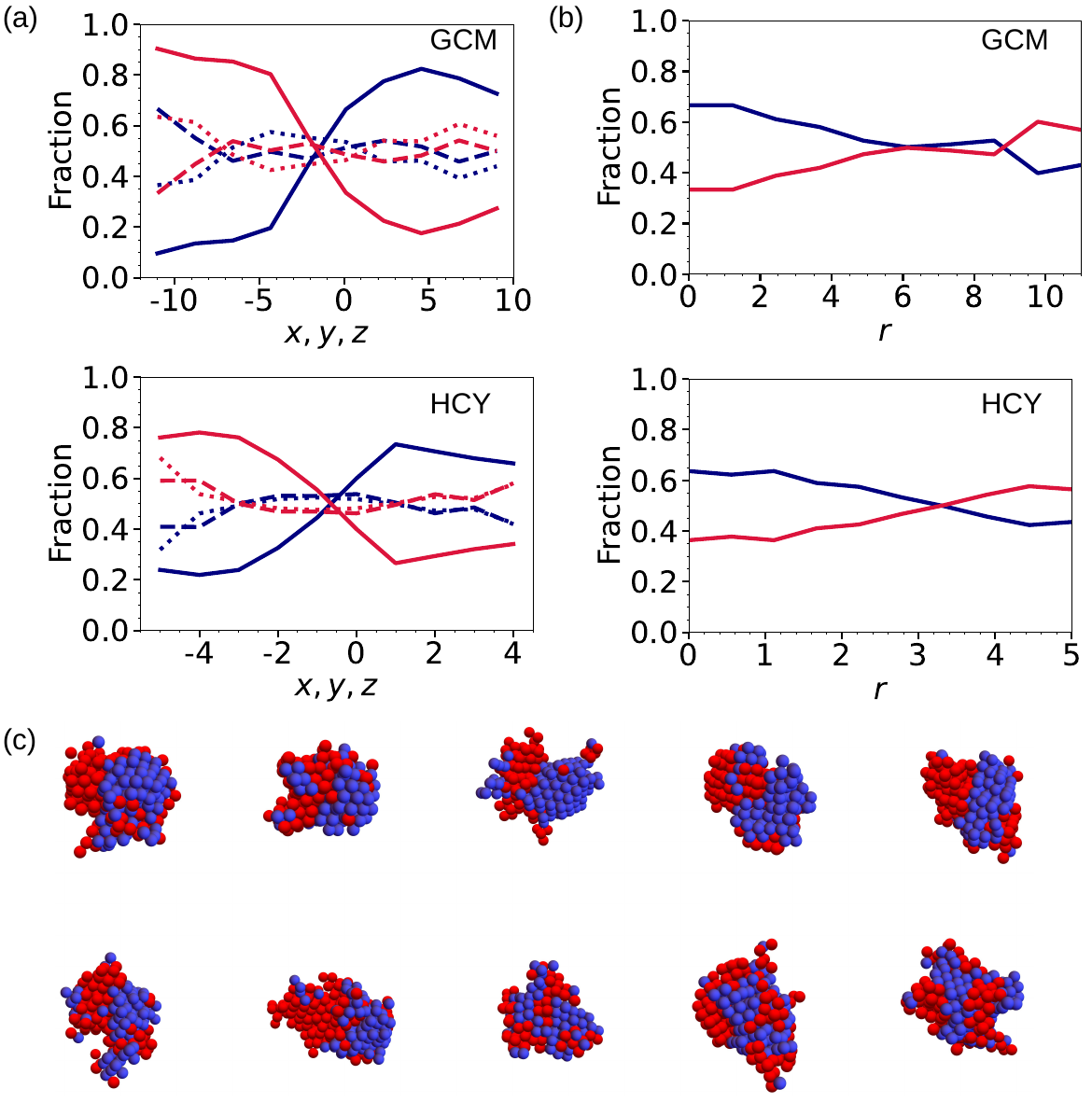}
    \caption{(a) The composition of the critical solid-like cluster at SP2 for the GCM (top) and HCY (bottom) systems. The critical cluster is aligned in such a way that the $x$-axis lies in the direction of the maximum variation of the BCC-like particles. The solid red and blue lines represent the variation of the fraction of the BCC- and FCC-like particles along $x$-axis. The dashed and dotted lines represent the same along the $y$- and z-axes. Here, we have chosen only those trajectories for which the $f_{\rm bcc}$ in the critical cluster lies between $0.48$ and $0.52$. (b) Composition profiles (number of particles of the $i^{\rm th}$ solid phase in a shell of radii $r$ and $r + \Delta r$ divided by the total number of solid-like particles in that shell) for the FCC-like (blue solid line) and BCC-like (red solid line) particles in the critical cluster as a function of the distance ($r$) from the center of the cluster for the GCM and HCY systems. (c) The representative snapshots of the critical solid-like cluster for different independent trajectories with $f_{\rm bcc} \sim 0.5$ for the GCM (top) and HCY (bottom) systems. It is evident from the figures that the critical cluster does not contain a core-shell-like structure where one type of solid polymorph in the core is wetted by the other type of polymorph.}
    \label{fig3x}
\end{figure}

To understand the detailed microscopic arrangement of particles of different polymorphic identities in a growing solid-like cluster, we have further investigated the composition of the critical clusters in the vicinity of the triple point (\textit{i.e}, at SP2) where a non-classical transition pathways via the participation of the intermediate metastable phase is a possible scenario (for example, see Ref.~\emph{\citenum{iwamatsu_2011, bagchi_osr_2013, poole_1}}). This participation of the intermediate phase often gives rise to a core-shell like structure of the critical cluster, as discussed previously. To probe the precise phase transition pathways, we calculated here the composition profiles of the critical solid-like cluster averaged over $\sim 10-20$ independent successful trajectories, along with the representative snapshots of the critical cluster. The trajectories chosen here are those in which the fraction of BCC-like particles in the critical cluster is in the range of $0.48 - 0.52$. In Fig.~\ref{fig3x}a we show the composition of the critical cluster for both model systems. To compute these composition profiles, the critical cluster is aligned in such a way that the $x$ axis lies in the direction of the maximum variation of the BCC-like particles. As we move away from the center of the cluster, along the positive $x$ direction the FCC-like particles increase, while in the opposite (negative $x$) direction the BCC-like particles increase. The composition of the FCC and BCC-like particles along the other ($y$ and $z$) directions remains almost the same. This picture differs markedly from the core-shell structure of the critical cluster reported in model lattice systems and in c-DFT-based studies in the vicinity of the triple point~\cite{poole_1, poole_2, bagchi_osr_2013}. 

We further computed the composition profiles (number of particles of the $i^{\rm th}$ solid phase in a shell of radii $r$ and $r + \Delta r$ divided by the total number of solid-like particles in that shell) for the FCC-like and BCC-like particles in the critical cluster as a function of the distance from the center of the cluster (see Fig.~\ref{fig3x}b). We again find no signatures of the core-shell structure of the critical cluster (for the core-shell structure, one would expect a significant dominance of one type of polymorph over the other near the core of the cluster and opposite near the surface). These observations are further supported by representative snapshots of the critical cluster for different independent trajectories with an approximate fraction of BCC-like particles of $0.5$ (Fig.~\ref{fig3x}c). It is evident from the snapshots that the growing critical clusters show an interpenetrating arrangement of FCC- and BCC-rich sub-clusters rather than a well-defined core-shell-like structure often reported in theoretical and computational studies (especially on lattice models) exhibiting similar free energy landscapes (for example, see Refs.~[\emph{\citenum{iwamatsu_2011, bagchi_osr_2013, poole_1, poole_2}}]). We also note that the final crystals formed after the phase transition at different state points ranging between SP1 and SP3 exhibit a similar interpenetrating arrangement of FCC and BCC-like particles near SP2 (see Fig.~S3 of the Supplementary Material). 
\subsubsection{Plausible origin of the observed deviation from the core-shell-like nucleation scenario}
\textbf{Kinetic origin:} In the conventional core–shell picture, local fluctuations resemble the phase with the lowest surface free energy relative to the surrounding fluid, which may not necessarily be the globally stable phase. Growth in this model occurs in two steps: diffusion of particles from the metastable fluid and their subsequent rearrangement within the shell composed of the intermediate metastable phase (in our case, the BCC phase)~\cite{iwamatsu_2017}. For the case where the intermediate metastable phase is a solid (unlike in protein crystallization or colloids interacting via short-range interactions, where the crystal nucleates from a dense fluid), particle diffusion is significantly faster in the fluid than in the metastable intermediate solid. For the phase transition to the globally stable solid phase, the rate-limiting step would be diffusion within the shell formed by the intermediate metastable phase (assuming the core-shell scenario). As a result of this kinetic bottleneck, the fluid may fully transform into the intermediate BCC phase before the thermodynamically most stable FCC phase has a chance to nucleate within the BCC on simulation time scales (note that structure formation requires a collective many-particle rearrangement). Therefore, although the core–shell pathway may be thermodynamically favorable, it may not be realized in off-lattice models, where particle transport in space governs the nucleation and subsequent growth of clusters.

Furthermore, if the nucleation barriers for the BCC and FCC phases are comparable, the FCC phase may also begin to grow, either homogeneously or heterogeneously (at the interfaces of growing intermediate metastable BCC-like crystallites). This would lead to competitive growth of the BCC and FCC phases, resulting in cluster morphologies where the BCC and FCC-like particles are intercalated or stacked (see Fig.~\ref{fig3x}b) --- rather than a sequential core–shell-like mechanism. Thus, in the vicinity of the triple point (where three distinct phases exit), depending on the free energy landscape (i.e., the relative stability of the different phases with respect to the metastable fluid) and the intrinsic dynamics of particles within the various concerned phases, diverse nucleation pathways can emerge, potentially deviating from the core–shell route. These findings suggest that theoretical models assuming core–shell formation or instantaneous diffusion equilibria at interfaces (FCC–fluid, BCC–fluid, and FCC–BCC interfaces maintaining the respective equilibrium concentrations) may need to explicitly account for kinetic (particle transport) bottlenecks and the possibility of heterogeneous nucleation.

\textbf{Thermodynamic origin:} From a thermodynamic perspective, the sequential core-shell and competitive growth mechanisms differ fundamentally. In crystallization, a core-shell structure forms when the most stable phase does not directly interface with the metastable fluid, instead growing around an intermediate metastable phase. In contrast, competitive growth involves both stable phases growing in direct contact with the metastable fluid. If the interfacial surface tension between the fluid and the stable solid is significantly higher than that between the fluid and an intermediate metastable solid, a core-shell structure is favored. In this configuration, the growing cluster minimizes its interfacial free energy by encapsulating the stable phase within the metastable one, effectively ``hiding" the stable core inside a shell of the intermediate phase. A similar phenomenon, though not directly linked, occurs in the case of hydrophobic collapse during protein folding, where the hydrophobic part of a protein positions itself at the core of the native structure to avoid contact with water. However, if the surface tension of the fluid-stable solid interface is comparable to that of the fluid-intermediate metastable solid interface, it is expected to have competitive growth of the stable phases.  

Additionally, for wetting-mediated (or core-shell-like) growth, the stable polymorph needs to be significantly stable with respect to the intermediate metastable phase to enable the formation of a stable phase cluster within the intermediate metastable phase fluctuations on accessible computational time scales. If the stable polymorph is only marginally more stable than the intermediate metastable polymorph --- as is the case at SP2 in the present study --- the thermodynamic driving force for forming a core of the stable phase within the growing intermediate metastable polymorph (BCC in our case) will be insufficient.
 In this situation, one would rather expect either the formation of a single-phase cluster of the phase with lower surface tension or a competitive growth of both polymorphs if their interfacial tensions with the parent metastable fluid phase are comparable, as discussed above. 

The above two hypothesized scenarios need careful computational validation and could be an interesting avenue for future research. In the next section, we have developed a machine learning approach to unravel the hidden signatures of polymorph selection and nucleation pathways (discussed above) in the local structures of the metastable fluid phase.  
\subsection{Polymorph selection information encoded in the structure of the metastable fluid: a machine learning approach} \label{subsec_3}
In our machine learning approach, we have used a structural descriptor based on persistence homology (PH; a topological data analysis method) to capture the local structure heterogeneity in a metastable fluid phase. This choice is inspired by recent studies that demonstrate the utility of PH in uncovering hidden local or medium-range structural features in disordered systems, especially in the context of amorphous solids and amorphous-to-amorphous phase transitions~\cite{nakamura2015persistent, hong2019medium, hiraoka2016hierarchical, gadha_jcp_2024} where traditional bond-orientational order parameters~\cite{, lechner_jcp_2008} were found to be inadequate in capturing hidden local structural heterogeneities. 
\begin{figure}
    \centering    \includegraphics[width=\linewidth]{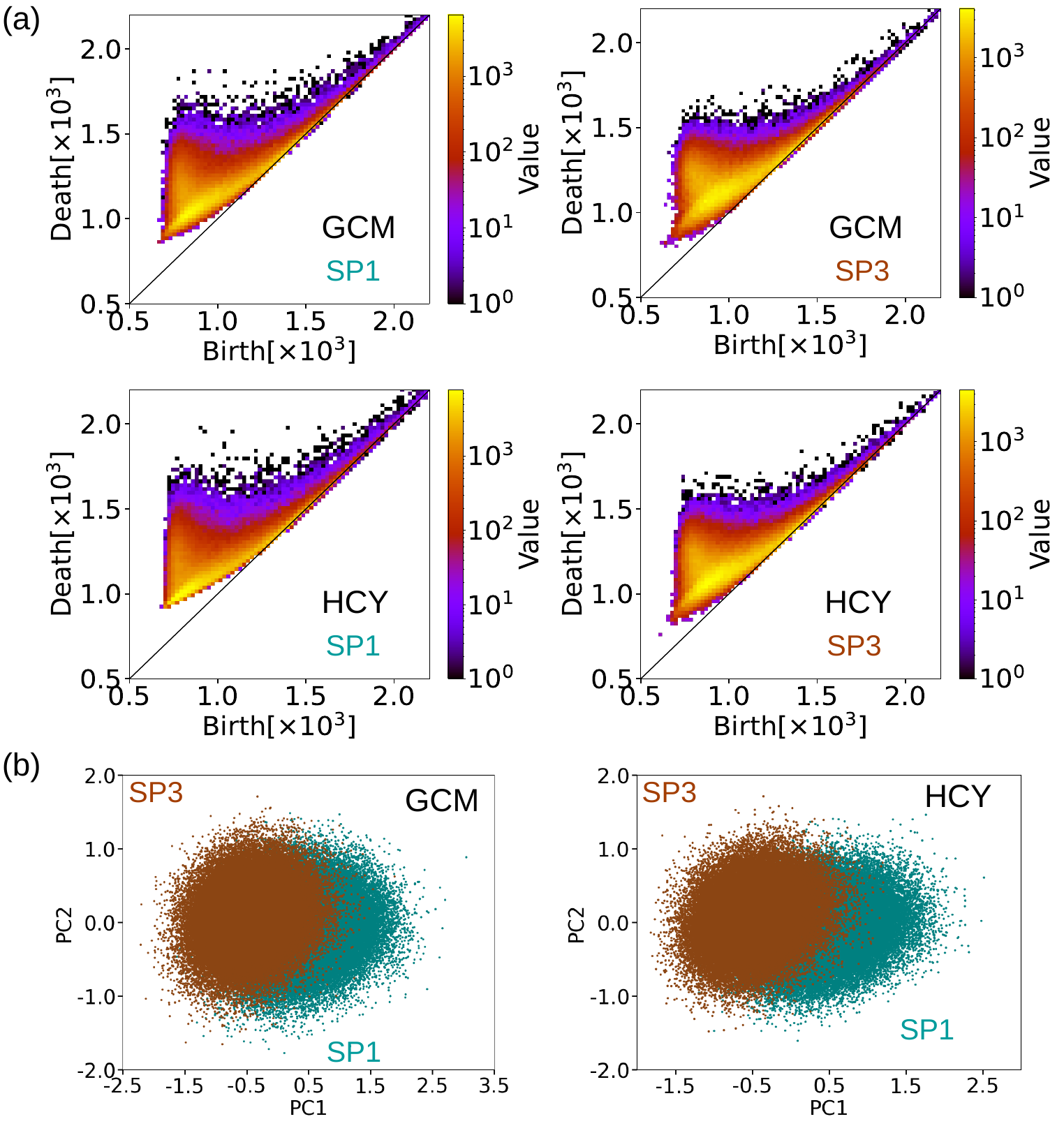}
    \caption{(a) We show the local persistence diagram (PD) of the GCM and HCY systems at SP1 and SP3. Here, we have chosen $32$ neighbors of a tagged central particle to define the local environment for both GCM and HCY systems. The local PD of the system given here is constructed by combining the local PD of all the particles in the system. (b) The results of the principal component analysis (PCA) of local PDs are shown for both GCM and HCY systems. The teal color represents SP1, and the brown color represents SP3. It is evident that the structural differences between these two state points are not fully resolved using two principal components (PC1 and PC2) for both model systems. This indicates the necessity of incorporating additional principal components to effectively capture the hidden structural distinctions between the two state points.}
    \label{fig_pd}
\end{figure}
\begin{figure}
    \centering
    \includegraphics[width=0.98\linewidth]{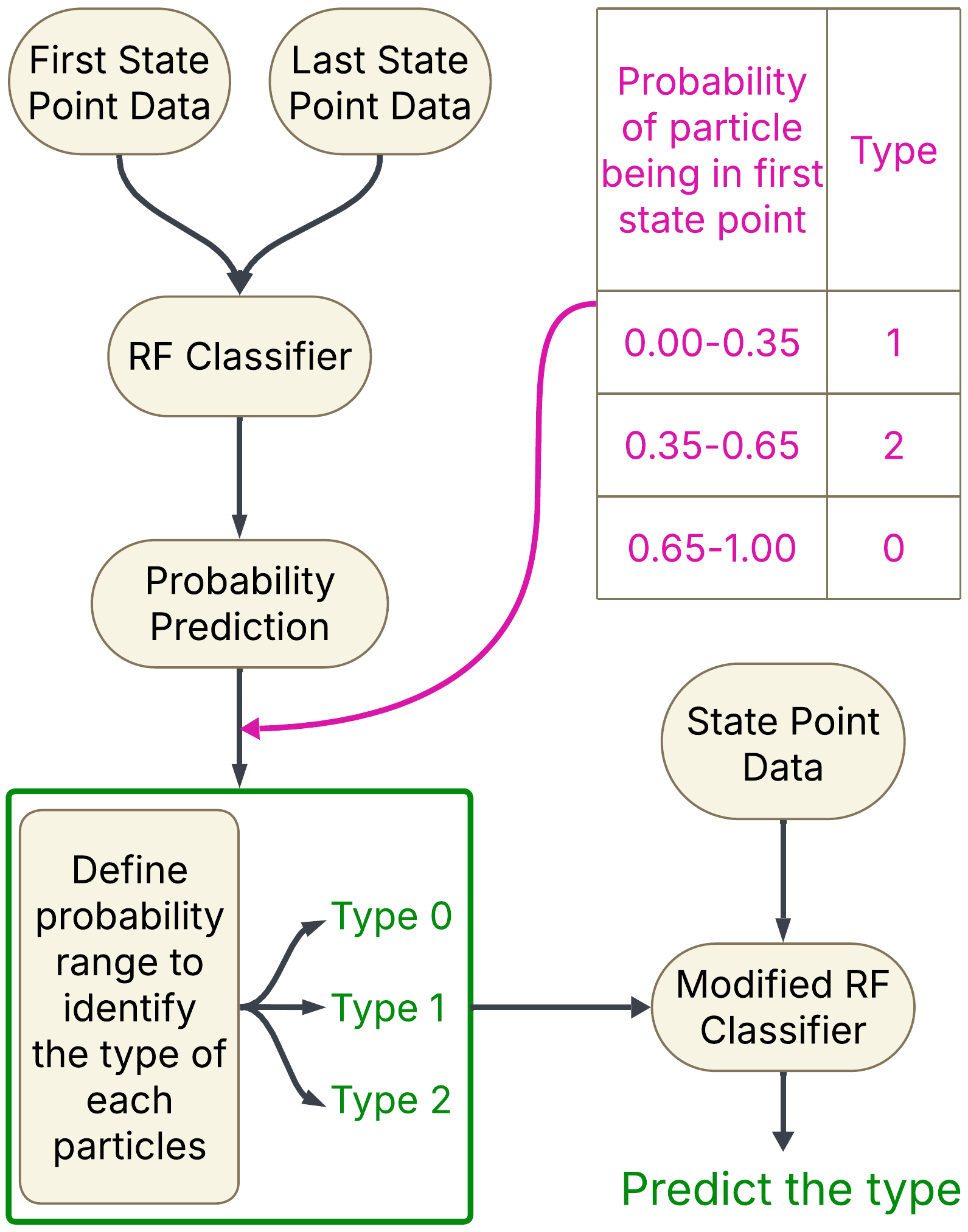}
    \caption{The RF-SLOP algorithm is represented. First, the information from the local PDs of each particle at the extreme state points (SP1 and SP3) is labeled and fed into the RF classifier model. The RF model classifies each particle into one of the two classes --- SP1 (`$0$') or SP3 (`$1$') --- based on the predicted probability. To capture any uncertainties in classification, a third category (`$2$') can be defined for particles with predicted probabilities in the intermediate range, typically between $0.35-0.65$. These are identified as potentially misclassified particles. With this modified three-class scheme (modified RF classifier), other state points are projected onto this classification, and the probability of each particle belonging to each class is then computed.}
    \label{fig_rf}
\end{figure}
\subsubsection{Local topological distinction}
To probe the local structural topology from the PH analysis (see Section I in the Supplementary Material for details), we constructed local persistence diagrams (PDs)~\cite{becker2022, gadha_jcp_2024}. The system is represented by a point cloud consisting of particle coordinates within $[0,1]$, which was achieved by scaling the coordinates with the box length. This scaling helps us to separate out the true structural distinction between configurations from the ones arising from the change of density (or inter-particle spacing) on changing the thermodynamic conditions. To define the local environment, we considered a central particle and chose the nearest $n_{\rm p}$ neighbors. We then computed the PD of this $n_{\rm p}+1$ sized point cloud. In Fig.~\ref{fig_pd}a we report the local PD of the system (plotted for all particles) for the GCM and HCY systems at SP1 and SP3 for $n_{\rm p} = 32$. 

To quantify the structural (topological) differences between the metastable fluids at SP1 and SP3, we employed the principal component analysis (PCA) --- an unsupervised machine learning dimensionality reduction technique --- on the vectorized data set of the PDs and projected the result onto a reduced 2D space of PC1 and PC2 (see Section II of the Supplementary Material for details). This gives us two partly overlapping point clusters with $N$ points ($N$ is the number of particles in the system) in the PC1-PC2 plane corresponding to the fluid configuration at SP1 and SP3 for each model. In Fig.~\ref{fig_pd}b, we show the projection of $10$ independent configurations corresponding to each state point onto the PC1-PC2 plane. Our choice of $n_{\rm p}=32$ or higher indicates that we are well beyond the first shell and also incorporating particles from the second coordination shells, hence exploring the local topological features at medium-range scales. It is evident that the structural differences between the two state points are not fully resolved by the PCA for either of the two model systems. This indicates the necessity of incorporating additional principal components. We extended the analysis to include up to six principal components and found that simply adding more components does not significantly improve the accuracy of the results, and the distinction between the SP1 and SP3 state points remained unclear. Hence, we devised a supervised machine learning algorithm to effectively capture the hidden structural distinctions between the two state points.
\subsubsection{RF-SLOP algorithm} 
We have used the Random Forest (RF) algorithm --- a supervised machine learning technique used for tasks such as classification and regression --- on local PDs projected on a six-dimensional principal component plane. Initially, the labeled (SP1 or SP3) input data is divided into a training set and a test set. The training set is used to train the RF classifier, which is built from an ensemble of decision trees, with each tree independently determining the class to which the input data belongs. The final classification is determined by a majority vote of the decisions made by each of the decision trees. This final classification is called the predicted class (or label) of the RF model. After training the RF model with the training set, one can use the test set data to check the accuracy of predictions from the trained RF model. 

The detailed classification protocol goes as follows. The local PDs are constructed first for the extreme state points, SP1 and SP3, and PCA is applied on them and is labeled as `$0$' (or, FCC-like) for SP1, and `$1$' (or, BCC-like) for SP3. This data is used as training input for the RF classifier. Our training of the RF model predicted the probability of the test data belonging to class $0$. That is, high probability refers to input data belonging to the class $0$, and low probability means that input data might belong to class $1$. Based on the predicted probabilities from the RF model, a new class `$2$' is defined to account for misclassified particles within an intermediate probability range (typically $0.35–0.65$). A modified RF classifier is then constructed that incorporates this additional misclassified class. PDs for other state points are constructed and fed into this modified RF model for three-class classification. We refer to this method as ``RF-SLOP", short for Random Forest-Structural Local Order Parameter, and a schematic representation of the same is provided in Fig.~\ref{fig_rf}. 
\begin{figure}
    \centering
    \includegraphics[width=0.97\linewidth]{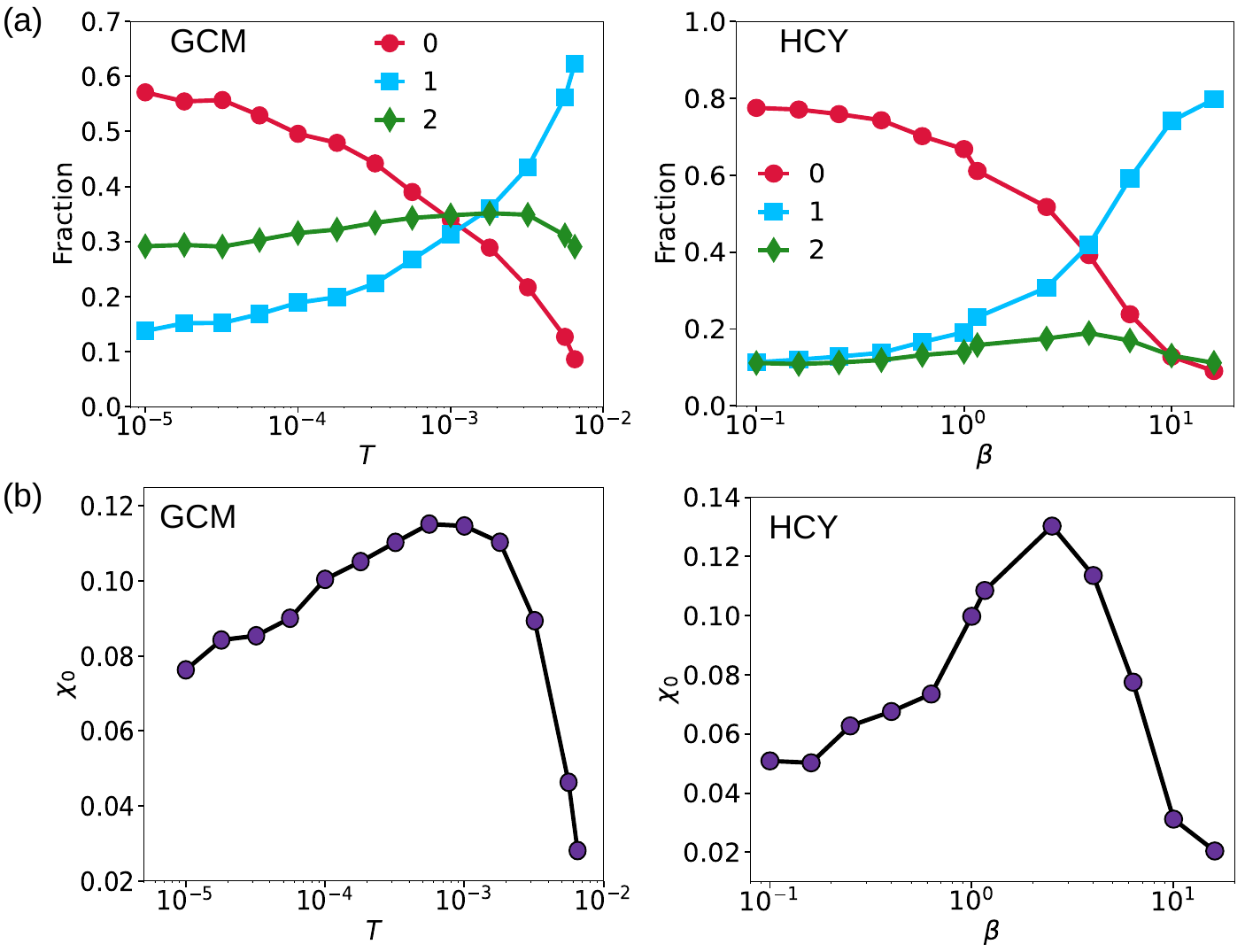}
    \caption{(a) The fractions of particles predicted to belong to class `$0$' (red), class `$1$' (blue), and class `$2$' (green) are shown as function of $T$ for the GCM, and $\beta$ for the HCY systems. As we move from SP1 to SP3 for both models, a decrease in the fraction of class `$0$' particles and a corresponding increase in class `$1$' particles is observed. Additionally, a slight increase in the fraction of class `$2$' (misclassified) particles appears near the crossover between class `$0$' and class `$1$'. (b) The fluctuations in the population of the RF-SLOP predicted class `$0$' are shown for both the model systems. A maximum in the $T$ (or, $\beta$ for the HCY system)-dependence of the population fluctuation is observed near the temperatures where Monte Carlo simulation data indicate an enhancement in the compositional heterogeneity during the phase transition (see Figs.~\ref{fig2} and~\ref{fig3}).}
    \label{fig_ml}
\end{figure} 
\subsubsection{RF-SLOP results}
Figure \ref{fig_ml}a shows the fraction of particles belonging to class `$0$', `$1$' or `$2$' as predicted by the RF-SLOP method with $n_{\rm p}=32$ at different state points reported in Table~\ref{tab1} for both GCM and HCY systems. All particles in each state point were initially labeled as `$0$’ as a starting point, since we employed a supervised learning scheme that requires labeled input data for training. When these labeled data were fed into the RF-SLOP model, it predicted the probability that each particle belongs to any of the three predefined classes, `$0$', `$1$', or `$2$'. As we move from SP1 (where the FCC phase is globally stable) to SP3 (where the BCC phase becomes globally stable), the fraction of particles classified as `$0$' decreases while that of particles classified as `$1$' increases. A slight increase in particles predicted to be in class `$2$' (or, misclassified particles) can be detected near the temperature where we observe a crossover from `$0$' to `$1$' dominance. That is, as one moves from SP1 to SP3, the fraction of particles identified as FCC-like decreases, while that of BCC-like particles increases. During this transition, a significant number of particles cannot be clearly classified as either FCC-like or BCC-like. This misclassified particle fraction is more for the GCM than the HCY. Detecting such polymorphic identity selection signatures, hidden within the metastable fluid phase, using local geometrical order parameters has been extremely challenging even for systems like hard spheres~\cite{tanaka_2012, tanaka_soft_2012, tanaka_pnas, marjolin_2023}. To check the sensitivity of these results on the choice of $n_{\rm p}$, we performed RF-SLOP for $n_{\rm p} = 40$ and found that the results are not remarkably sensitive to the choice of $n_{\rm p}$ (see Fig.~S4 in the Supplementary Material). It is important to note that if the analysis had been initiated with an initial label of `$1$’ instead, the results would remain unchanged. This means that the initial label serves purely as a placeholder, enabling the supervised learning algorithm to begin the classification process.

We also calculated fluctuations in the population (defined in same way as in Fig.~\ref{fig3}) of class `$0$' particles in various independent trajectories, as shown in Fig.~\ref{fig_ml}b for both the GCM and HCY systems. Interestingly, we find that the fluctuations exhibit a maximum for both models near the temperatures where the Monte Carlo simulation data show enhanced compositional diversity. This observation is quite remarkable, especially considering that during the training of our model, only the extreme state points were used, where no signs of competitive nucleation or significant compositional heterogeneity is present.

These findings suggest that the PH-based local structural descriptor has strong potential for revealing important signatures of polymorph selection and phase transition pathways hidden within metastable fluids. One can also explore other structural descriptors, such as smooth overlap of atomic positions (SOAP)~\cite{soap_1, soap_2, soap_3} and various variants of the bond-orientational order parameters (BOPs)~\cite{bop, lechner_jcp_2008, filion_prl_2021}, as we do not rule out the possibility that modifications of the traditional order parameters (such as, BOPs) could achieve even greater robustness.
\section{\label{sec_4}Conclusions}
In this study, we have employed MC simulations on prototypical colloidal model systems, belonging to two different classes, soft and hard core, and mimicked via the GCM and HCY potentials, to elucidate the thermodynamic control of polymorph selection during crystallization. By systematically perturbing the free energy landscape --- rendering the FCC phase stability from being globally stable to metastable with respect to the BCC phase --- we observe a polymorphic transition from FCC-dominated to BCC-dominated nucleation. This crossover proceeds through an intermediate regime wherein both polymorphs emerge either selectively or competitively, accompanied by pronounced polymorphic composition fluctuations reminiscent of near-critical phenomena. Detailed structural analysis of critical solid-like clusters, particularly in the vicinity of the triple point --- where fluid, FCC, and BCC phases coexist --- reveals a spatially interpenetrated arrangement of FCC- and BCC-like motifs, distinct from the two-step (core-shell-like) nucleation pathway mediated via metastable crystalline intermediate(s). We have further discussed the plausible reasons behind the deviation of the phase transition pathways from the expected core-shell-like scenario. 

Furthermore, to uncover hidden signatures of polymorphic bias and phase transition pathways encoded in the metastable fluid, we employ machine learning methodologies incorporating structural descriptors derived from PH, a topological data analysis technique sensitive to many-body structural correlations and local geometric features. This approach reveals the signatures of polymorph selection in the metastable fluid structure as well as the enhanced composition fluctuations in the vicinity of the triple point (i.e., near SP2). 

We believe that the findings of this work offer deeper insights into the rich phase transition pathways in systems characterized by complex free energy landscapes. These findings may also have important implications for the rational design of materials where the selection of a specific polymorph is crucial (for example, in pharmaceuticals, photonic materials, and energy applications).

\begin{acknowledgments}
R.S.S. acknowledges financial support from DST-SERB (Grant No. CRG/2023/002975). G.R. acknowledges financial support from IISER Tirupati. M. S. acknowledges financial support from DST-SERB (Grant No. SRG/2020/001385).  A.K. acknowledges financial support from IIT Goa. The computations were performed at the IISER Tirupati computing facility and at IIT Goa. 
\end{acknowledgments}

\bibliographystyle{apsrev4-2}
\bibliography{main_final}
\pagebreak
\widetext
\begin{center}
\textbf{\large Supplementary Material}
\end{center}
\setcounter{equation}{0}
\setcounter{figure}{0}
\setcounter{table}{0}
\setcounter{page}{1}
\setcounter{section}{0}
\makeatletter
\renewcommand{\theequation}{S\arabic{equation}}
\renewcommand{\thefigure}{S\arabic{figure}}
\begin{figure*}[htbp!]
    \centering
    \includegraphics[width=0.5\linewidth]{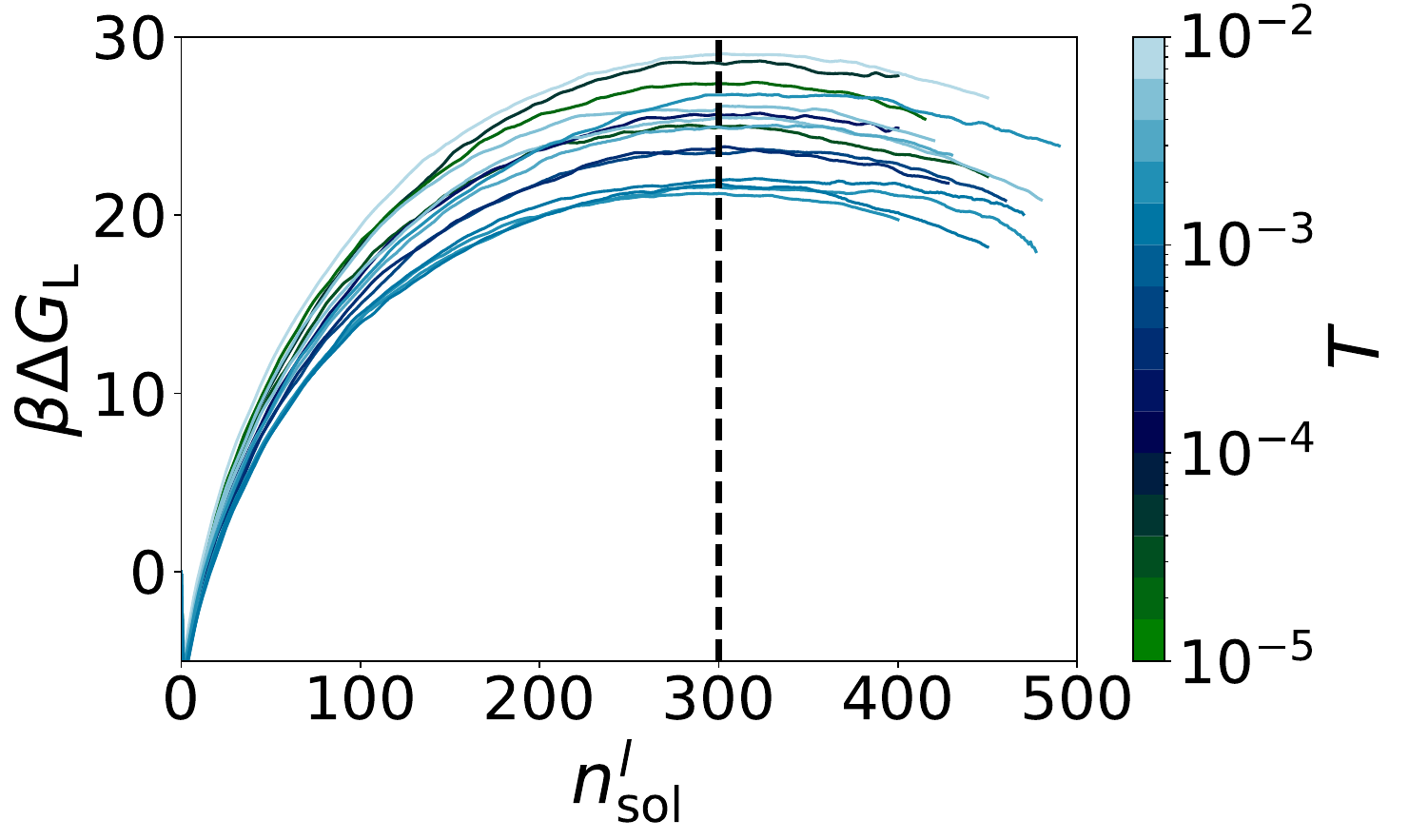}
    \caption{The nucleation free energy profile for the largest solid-like cluster ($n_{\rm sol}^l$) at different state points for the GCM model (see Table I in the main text).}
    \label{figsm1}
\end{figure*}

\begin{figure*}[htbp!]
    \centering
    \includegraphics[width=0.6\linewidth]{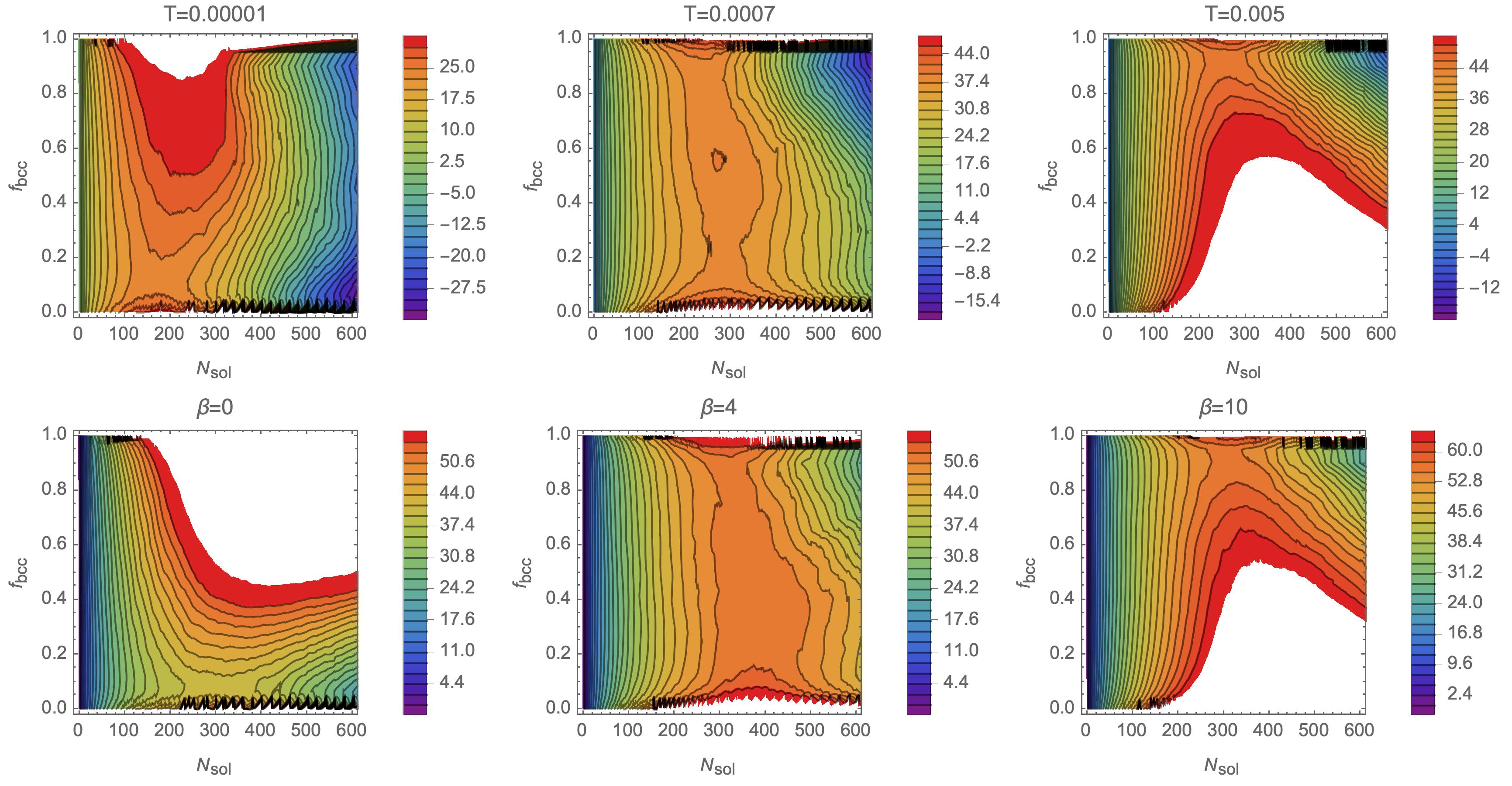}
    \caption{ The computed two-dimensional (2D) free energy surface (FES) is shown as a function of the fraction of BCC-like particles ($f_{\rm bcc}$) and the total number of solid-like particles ($n_{\rm sol}$) in the system for the GCM (top) and HCY (bottom) model systems. We selected three representative state points: (left) the FCC phase is the globally stable phase (SP1), (middle) near the triple point where the metastable fluid coexists with marginally stable FCC and BCC phases, and FCC is globally stable (SP2), and (right) the BCC phase is globally stable (SP3). To reduce computational cost, we used a smaller system size of $N = 3000$ for the these calculations, and accordingly adjusted the state points to cater to the finite-size effects. We note a flattening of the FES for the SP2, consistent with the presence of enhanced composition fluctuations during the phase transition. This observation aligns with the results shown in Figs.~2 and~3 of the main text.}
    \label{figsm3}
\end{figure*}

\begin{figure*}[htbp!]
    \centering
    \includegraphics[width=0.6\linewidth]{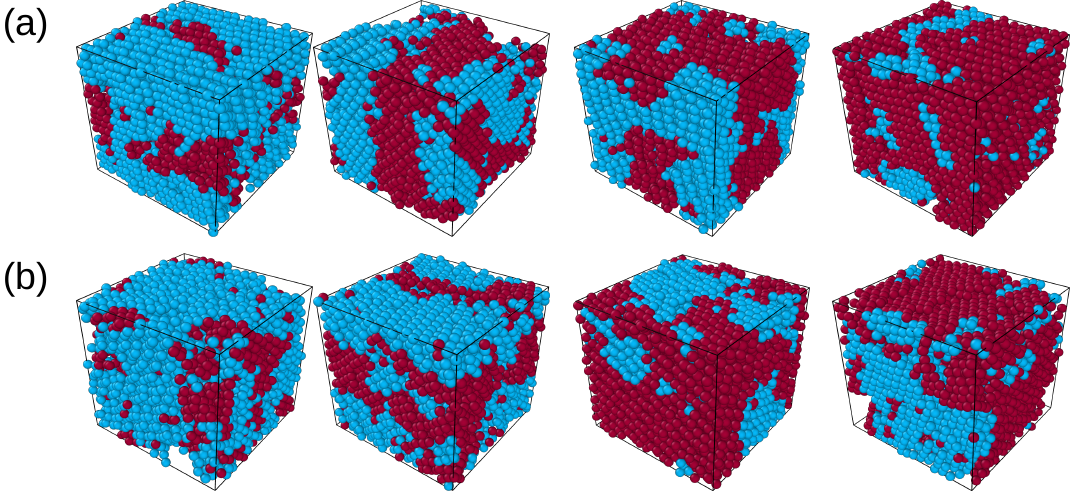}
    \caption{(a) The snapshots of the final solid configuration with $f_{\rm bcc}=0.2,0.5$, $0.6$, and $0.8$ for the GCM (a) and HCY (b) model systems. The FCC-like particles are represented in blue and the BCC-like in red.}
    \label{figsm4}
\end{figure*}

\begin{figure*}[htbp!]
    \centering
    \includegraphics[width=0.6\linewidth]{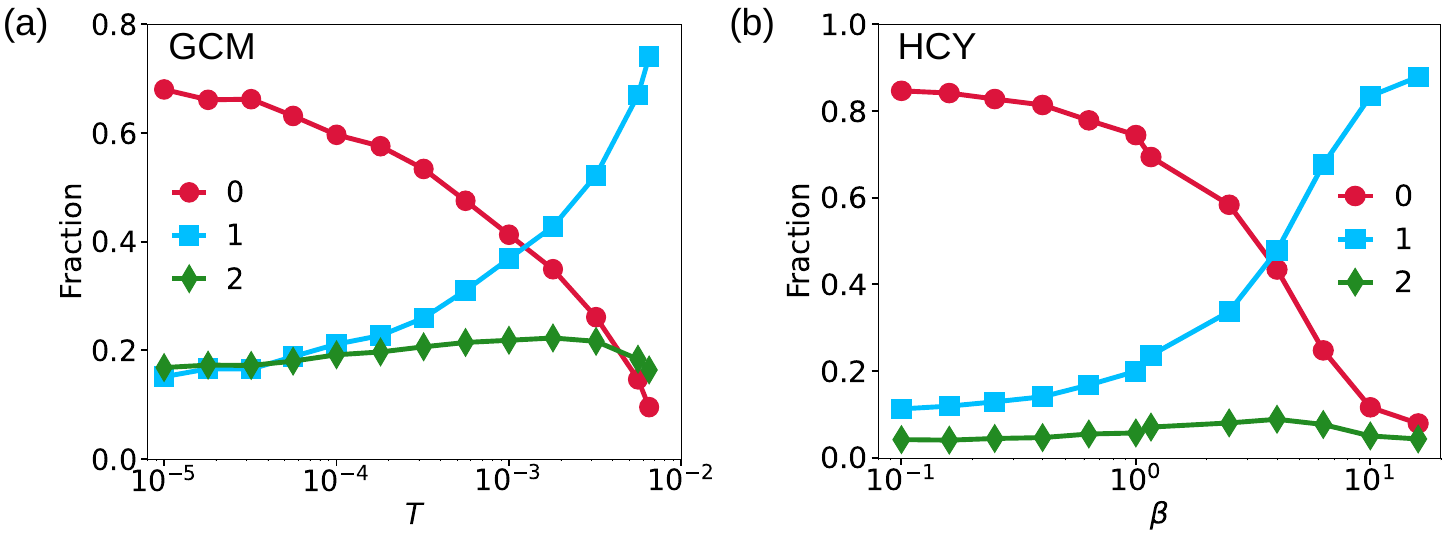}
    \caption{The results of the RF-SLOP method for $40$ nearest neighbors for GCM and HCY model systems are shown. It is evident that the overall trend remains similar to the one shown in Fig.~7a of the main text for $32$ nearest neighbors.}
    \label{figsm1}
\end{figure*}

\section{Persistence Homology (PH) Computation}
The PH computing mechanism is as follows: (a) Consider an atomic configuration of $N$ atoms with coordinates, $Q=(\mathbf{x}_1,\mathbf{x}_2,..,\mathbf{x}_N)$ with input radii $R=(r_1,r_2,...,r_N)$, which is represented in the form of a point cloud where the coordinates are scaled between $0$ and $1$. To compute PH, we place a spherical ball of radius $r_j$ centered at $\mathbf{x}_j$, and increase the radius of the ball as $\sqrt{r_j^2 + \varepsilon}$ with increasing $\varepsilon$, the scale parameter in the calculation. Line segments are introduced to connect the corresponding atoms as these balls intersect. Additional line segments emerge with a further increase in $\varepsilon$, eventually forming a ring as they connect multiple atoms end-to-end. The moment that this ring, say $X$, first appears is called its birth, and the corresponding $\varepsilon$ is recorded as the birth scale $b_X$. As the value of $\varepsilon$ increases further, the expansion of the balls causes them to overlap and penetrate one another, leading to the disappearance of the ring $X$. The scale at which this disappearance occurs is recorded as the death scale $d_X$. With varying $\varepsilon$ values, multiple rings may form (birth) and later vanish (death). The resulting diagram, which is the collection of all pairs of birth and death of these rings, is known as a persistence diagram (PD). In a three-dimensional (3D) system, connections between the balls can form cavities or voids, in addition to rings. A one-dimensional PD (named PD1) is a persistence diagram constructed exclusively from ring information, while a two-dimensional PD (named PD2) is built from cavity information. Note that in a PD, all points consistently lie above the diagonal line, where this (diagonal) line represents equal birth and death values. In this work, our focus is on analyzing only PD1, and therefore, we will refer to it simply as PD. Each PD consists of three components: the birth scale plotted on the x-axis, the death scale plotted on the y-axis, and the color representing the density of points (in this case, rings). We have used HomCloud~\cite{obayashi2022persistent} as well as the Ripser package implemented for Python~\cite{ctralie2018ripser} to compute the PDs reported in this work. 

\section{Persistence diagram (PD) analysis}
For each tagged particle, a point cloud of size $n_{\rm p} + 1$ was constructed, where $n_{\rm p}$ is the fixed number of nearest neighbors considered. The corresponding persistence diagrams (PDs) were generated using Vietoris–Rips filtration. (Alpha filtration was also tested, and the results were found to be insensitive to the choice of filtration method.) Each PD was then divided into a $48 \times 48$ grid—further increasing the resolution did not significantly affect the results. Subsequently, the persistence image technique was applied to vectorize each PD into a one-dimensional array of length $48 \times 48 = 2304$. Thus, a total of $N$ PDs gives a 2D array of size $[N \times 2304]$, where $N$ is the system size. This matrix was used as input for principal component analysis (PCA), which reduced the dimensionality to $[N \times 2]$. For each state point, $10 \times N$ PDs (denoted as $N_{\rm all}$) were obtained from $10$ independent simulations. PCA was performed on the full $[2N_{\rm all} \times 2304]$ dataset, resulting in a reduced representation of size $[2N_{\rm all} \times 2]$. It is important to note that this analysis focuses exclusively on one-dimensional homology (PD1), which captures information about ring-like topological features.

\end{document}